\documentclass{article}
\usepackage{emulateapj,epsfig,apjfonts}
\long\def\***#1{{\scshape ***#1***}}

\makeatletter

\newenvironment{inlinefigure}{%
\def\@captype{figure}%
\noindent\begin{minipage}{0.999\linewidth}\begin{center}}
{\end{center}\end{minipage}\smallskip}
\makeatother

\lefthead{HU, M{\sc{c}}MAHON, \& COWIE}
\righthead{AN EXTREMELY LUMINOUS GALAXY AT $z$ = 5.74}
\newcommand{\msun}{$M_{\odot}$}
\slugcomment{To appear in The Astrophysical Journal (Letters)}

\begin{document}
\title{An Extremely Luminous Galaxy at \lowercase{$z$} = 5.74}
\author{Esther M. Hu\altaffilmark{1,2},
Richard G. McMahon\altaffilmark{1,3}}
\and 
\author{Lennox L. Cowie\altaffilmark{1,2}}
\altaffiltext{1}{Visiting Astronomer, W. M. Keck Observatory,
  jointly operated by the California Institute of Technology, the University
  of California, and the National Aeronautics and Space Administration.}
\altaffiltext{2}{Institute for Astronomy, University of Hawaii, 2680 Woodlawn
  Drive, Honolulu, HI 96822}
\altaffiltext{3}{Institute of Astronomy, University of Cambridge, 
  Madingley Road, Cambridge CB3\thinspace{0HA}}

\begin{abstract}
We report the discovery of an extremely luminous galaxy lying at a redshift
of $z=5.74$, SSA22-HCM1.  The object was found in narrowband imaging of the
SSA22 field using a 105~\AA\ bandpass filter centered at 8185~\AA\ during the
course of the Hawaii narrowband survey using LRIS on the 10 m Keck\,II
Telescope, and was identified by the equivalent width of the emission
($W_{\lambda}$(observed)=175 \AA, flux = $1.7 \times 10^{-17}$ erg cm$^{-2}$
s$^{-1}$).  Comparison with broadband colors shows the presence of an
extremely strong break ($> 4.2$ at the 2$\sigma$ level) between the $Z$ band
above the line, where the AB magnitude is 25.5, and the $R$ band below, where
the object is no longer visible at a $2 \sigma$ upper limit of 27.1 (AB
mags).  These properties are only consistent with this object's being a
high-$z$ Ly$\alpha$ emitter.  A $10\,800$ s spectrum obtained with LRIS yields
a redshift of 5.74.  The object is similar in its continuum shape, line
properties, and observed equivalent width to the $z=5.60$ galaxy, HDF 4-473.0,
as recently described by \markcite{weymann98}Weymann {et~al.}\ (1998), but is
2--3 times more luminous in the line and in the red continuum.  For $H_0 =
65\ {\rm km}\ {\rm s}^{-1}\ {\rm Mpc}^{-1}$ and $q_0 = (0.02, 0.5)$ we would
require star formation rates of around (40, 7) $M_{\odot}$ yr$^{-1}$ to
produce the UV continuum in the absence of extinction.
\end{abstract}


\section{Introduction}
Studies of high-redshift galaxies have now advanced well beyond $z=5$, with
the confirmation of a number of such systems using the Keck II 10-m telescope
over the past year \markcite{dey98, smitty2, weymann98}(Dey {et~al.}\ 1998;
Hu, Cowie, \& McMahon 1998; Weymann {et~al.}\ 1998).  A key issue for future
studies of distant, early star-forming galaxies is the incidence of bright,
high-redshift galaxies in the general field population.  Here we report on the
discovery of a luminous $z=5.74$ Ly$\alpha$-emitting galaxy in the Hawaii
Survey Field SSA22.

As part of our deep imaging searches to systematically identify and study
high-redshift galaxies in blank survey fields \markcite{smitty1,smitty2}(Cowie
\& Hu 1998; Hu {et~al.}\ 1998) and fields around high-redshift quasars
\markcite{br1202,br2237}(Hu, McMahon, \& Egami 1996; Hu \& McMahon 1996), we
have been conducting deep narrowband imaging on the Keck II 10-m telescope to
look for high-redshift Ly$\alpha$-emitting galaxies.  Previous results from
the deep LRIS narrowband surveys at 5390 \AA, or $z_{\rm Ly\alpha}\sim3.4$,
and 6741 \AA, or $z_{\rm Ly\alpha}\sim4.5$, have been reported in
\markcite{smitty1}Cowie \& Hu (1998) and \markcite{smitty2}Hu {et~al.}\ 
(1998).  The present object, which we will refer to as SSA22-HCM1, was
discovered during the course of a $z\sim5.7$ search using a 105 \AA\ bandpass
narrowband filter centered on 8185 \AA. The results of this survey will be
reported in detail elsewhere \markcite{smitty3}(Hu, Cowie, \& McMahon 1999).
We have chosen to describe this object separately because of its high
luminosity, which places it at the high end of the luminosity function for
optically selected galaxies at all redshifts.  If the light is produced by
massive star formation with a Salpeter mass function, the inferred rate is 40
$M_{\odot}$ yr$^{-1}$ $h_{65}^{-2}$ for $q_0=0.02$.

\section{Observations}

\begin{figure*}
\centerline{\epsfig{file=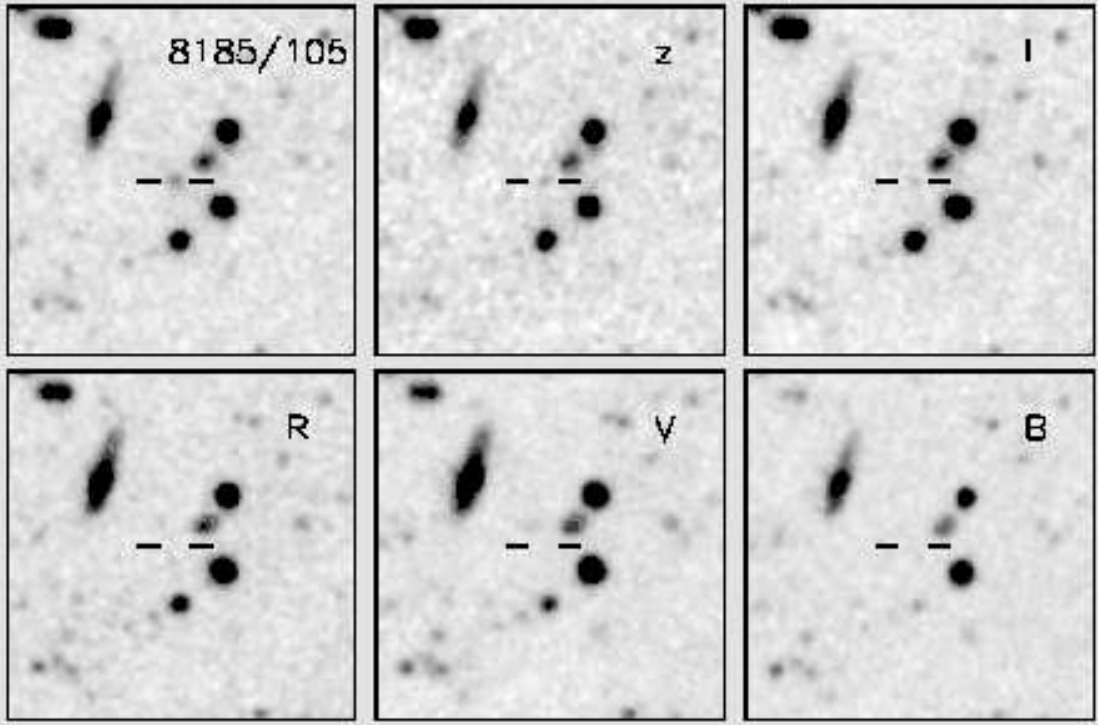,height=3.80in}}
\caption{\small Multicolor $B$, $V$, $R$, $I$, $Z$, and narrowband 8185/105
\AA\ images taken with LRIS on Keck of the $z=5.74$ object, SSA22-HCM1
(RA(1950): 22\ 15\ 05.97,  Dec(1950): $-$00\ 01\ 12.9).  Each panel is 30$''$
on a side.  The positions of the sources are marked, with a separation of
about $2\farcs5$ between tick marks.  The $I$ band image is weakly
contaminated by line emission, but most of the light arises from the
continuum, while the $Z$ band is free of emission.  The object is absent at
$B$, $V$, and $R$ in these extremely deep Keck LRIS images, which reach
$1\sigma$ limits of $B$=28.3, $V$=28.2, and $R=27.8$ for a $2''$ diameter
aperture (magnitudes in the AB system). $I(AB)$ for SSA22-HCM1 is 26.5 and
$Z(AB)$ is 25.5.}\label{fig:1}
\addtolength{\baselineskip}{-30pt}
\end{figure*}

Deep multi-color imaging data with LRIS \markcite{lris}(Oke {et~al.}\ 1995) on
the Hawaii Survey Field SSA22 have been described in \markcite{smitty1}Cowie
\& Hu (1998) and \markcite{smitty2}Hu {et~al.}\ (1998).  The $z=5.7$ Ly$\alpha$ 
survey of this region was carried out with narrowband exposures in the 8185
\AA\ filter obtained on UT 21 Aug 1998 as a sequence of 9 1200s exposures.
The present work augments the continuum data at the redder wavelengths by
doubling the $I$-band exposures (seven 360s exposures taken UT 21 Aug 1998),
and by obtaining a line-free red continuum through an RG850 filter, hereafter
referred to as $Z$ band, on UT 17 Sep 1998 as a dithered sequence of eighteen
330s exposures.  Conditions were photometric on these nights, with image FWHM
$\sim 0.65-0.7''$ for the $Z$ band exposures and $\sim 0.7-0.75''$ for the
$I$ and 8185/105 narrowband exposures. These data were calibrated with
spectrophotometric standards (Feige 110, Feige 15, and HZ4
[\markcite{massey88,massey90,oke90,stone96}Massey {et~al.}\ 1988; Massey \&
Gronwall 1990; Oke 1990; Stone 1996]) and Landolt standards
\markcite{landolt92}(Landolt 1992). The net exposure times for the combined
LRIS imaging in these bands are summarized in Table~\ref{tbl:1}, and
multi-color images of SSA22-HCM1 are shown in Fig.~\ref{fig:1}.

The $Z$ band is similar to the $z'$ filter used by the Sloan Digital Sky Survey
\markcite{frei94,fuk95,sloan_filt}(Frei \& Gunn 1994; Fukugita, Shimasaku, \&
Ichikawa 1995; Fukugita {et~al.}\ 1996) and to the Gunn $z$ \markcite{sgh83,
fuk95}(Schneider, Gunn, \& Hoessel 1983; Fukugita {et~al.}\ 1995) and Gunn $z_4$
\markcite{ssg89,fuk95}(Schneider, Schmidt, \& Gunn 1989; Fukugita {et~al.}\
1995) bands used in high-$z$ quasar searches. The combined filter + LRIS optics
+ CCD response curve yields an effective wavelength $\lambda_{\rm eff}=9271$
\AA\ (9166 \AA\ for a flat $f_{\nu}$ source) and FWHM=1430 \AA, with a peak
response at 8811 \AA.  The net response is less than 1\% of this maximum value
at 8185 \AA, so the $Z$ band provides a long-wavelength continuum measurement
uncontaminated by emission in the 8185 \AA\ narrowband filter.

\begin{inlinefigure}
\centerline{\epsfig{file=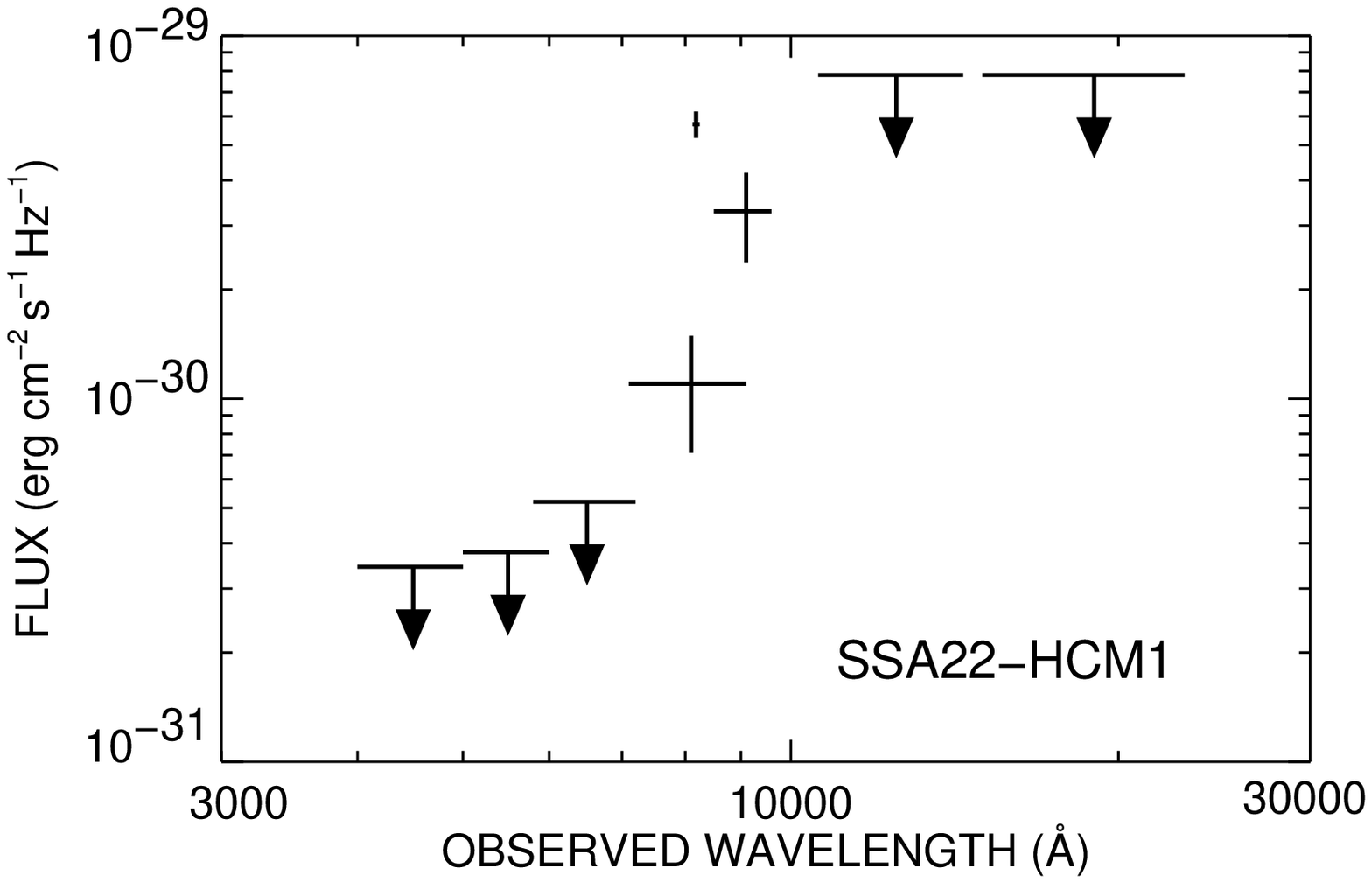,width=235pt}}
\vspace{-0.15in}
\caption{\small Spectral energy distribution for SSA22-HCM1 obtained using
LRIS, and showing the bandwidths and errors for measurements in $B$, $V$, $R$,
$I$, narrowband 8185/105 \AA\ (emission), RG850 (line-free continuum longwards
of the emission), $J$, and $H+K'$. Significantly, the object has no flux in
the $R$ band, where strong Lyman $\alpha$ forest absorption is expected to be
present. The $I$ band is slightly contaminated by Ly $\alpha$ emission, but
the $Z$ band measurement at longer wavelengths is a line-free continuum
measurement.  We use the $R$ and $Z$ bands to measure the discontinuity.  The
strong break ($2\sigma$ upper limit of 0.23 across the line) combined with the
strong emission line is a signature for Ly$\alpha$.}
\label{fig:2}
\end{inlinefigure}

\begin{inlinefigure}
\vspace{-0.9cm}
\centerline{\epsfig{file=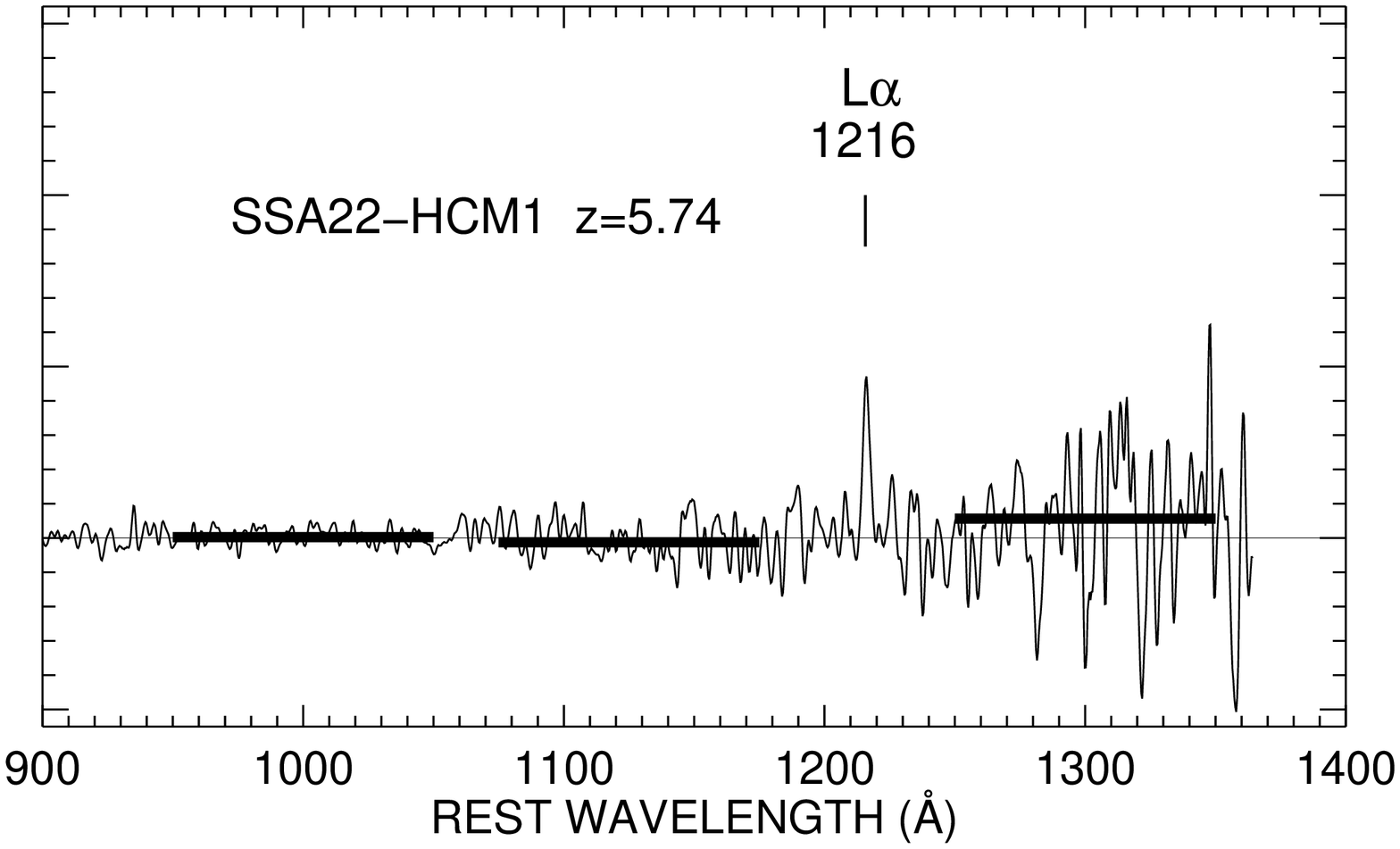,width=245pt}}
\vspace{-0.5cm}
\caption{\small Spectrum of the redshift $z=5.74$ galaxy SSA22-HCM1 obtained
in a 10,800 sec exposure using the $300\ell$/mm grating on  LRIS.  The heavy
lines show the averaged continuum values at the indicated rest-frame
wavelengths above and below the line.}
\label{fig:3}
\end{inlinefigure}

Candidate Ly$\alpha$ emitters were selected for spectroscopic followup from
sources in the 5$\sigma$ narrowband catalog with observed equivalent widths
greater than 80 \AA\ in the observed frame, which were also undetected in the
$V$ band down to the $2\ \sigma$ limit of 27.5.  For the current work,
photometric measurements are made over matched $2''$ diameter apertures, in
contrast to the $3''$ diameter apertures used in earlier work
\markcite{smitty1,smitty2}(Cowie \& Hu 1998; Hu {et~al.}\ 1998), and are
corrected to total magnitudes following \markcite{cowie_1}Cowie {et~al.}\
(1994).  The $1 \sigma$ error limits given in Table~\ref{tbl:1}, and in
\markcite{smitty1}Cowie \& Hu (1998) and \markcite{smitty2}Hu, Cowie, \&
McMahon (1998), are determined from flux measurements in random aperture
samples laid down at positions away from identified sources, since in the deep
exposures (cf.\ Fig.~\ref{fig:1}) magnitude limits are set by the background
faint source population.  These values can be up to $\sim0.7$ mags brighter
than limiting magnitudes estimated by background pixel statistics
\markcite{dey98}(e.g., Dey {et~al.}\ 1998), depending on color bandpass.

%
%
\begin{table*}
\tablenum{1}
{\small
\begin{center}
\centerline{\sc Table 1}
\vspace{1mm}
\centerline{\sc Photometric Data on the $z$=5.74 Galaxy SSA22-HCM1}
\vspace{2mm}
\begin{tabular}{lccccccccc}
\hline\hline
\noalign{\smallskip}
& {W$_{\lambda}$(obs'd)}   & {Flux}                            &
                           &                                   &
                           &                                   &
                           &                                   & \cr
& {(\AA)}                  & {(erg cm$^{-2}$ s$^{-1}$)}        &
{~$Z(AB)^{a}$}             & {$N(AB)^{b}$}                     &
{$I(AB)$}                  & {$R(AB)$}                         &
{$V(AB)$}                  & {$B(AB)$}           & {$K(AB)^{c}$} \cr
\hline
\noalign{\smallskip}
& 175 & $1.7\,(-17)$ &  25.5 & 24.5 & 26.6 & --28.1\ \ & --29.9\ \ & 29.8 & --
25.0\cr
{$\langle1\sigma\rangle$} & \phn20 & $2.2\,(-18)$ & 26.5 &
    27.2 & 27.3 & 27.8 & 28.2 &  28.3 & 24.9\cr
{$t_{exp}$ (s)} & &  & 5940 &  10800 & 5020 & 4320 & 5600 & 2440 & 14400\cr
\noalign{\smallskip}
\noalign{\hrule}
\noalign{\smallskip}
\multispan{10}{~~~$^a$ Schott RG850 filter, $\lambda_{\rm{eff}}=9271$ \AA,
  FWHM=1430 \AA\hfil}\cr
\noalign{\vspace{0.05cm}}
\multispan{10}{~~~$^b$ 8185/105 Narrowband filter\hfil}\cr
\noalign{\vspace{0.05cm}}
\multispan{10}{~~~$^c$ Upper limit, NIRC data taken under non-photometric
 conditions; $H+K'$ observations from UH 2.2m\hfil}\cr
\noalign{\vspace{0.05cm}}
\multispan{10}{~~~{\phantom{$^c$}} and CFHT of near comparable depth
 used as cross-check\hfil}\cr
\noalign{\vspace{-0.08cm}}
\end{tabular}
\end{center}
\vspace{-0.5cm}
\begin{center}
\begin{minipage}{15.3cm}
\parindent=2.5mm
{{\sc{Note.}---}$2''$ diameter aperture magnitudes.  A negative magnitude here
indicates that there is a negative flux in the aperture. The 1\ $\sigma$ limits
are derived from random aperture sampling off of sources in the composite Keck
exposures for each band ($t_{exp}$ given in seconds).\hfill\break}
\end{minipage}
\end{center}
\par
}
\vspace*{-0.85cm}
\label{tbl:1}
\end{table*}

Spectra of candidates were obtained on LRIS in multi-slit masks with the
300$\ell$/mm grating on the nights of UT 16 and 18 September 1998, and with the
400$\ell$/mm and 150$\ell$/mm gratings on the nights of UT 17 September and 22
October 1998.  Infrared images of SSA22-HCM1 at $K'$ were obtained with NIRC on
the Keck I telescope in 15000 s on UT 4 October 1998 under non-photometric
conditions.  These were cross-calibrated with $K'$ images of the SSA22 field of
near comparable depth from the Hawaii Survey program based on imaging with the
1024$^2$ UH IR camera, QUIRC, \markcite{quirc}(Hodapp {et~al.}\ 1996) at the UH
2.2m telescope and CFHT 3.6m telescope.  However, the $1\sigma$ detection
limits given in Table~\ref{tbl:1} are based on the QUIRC survey data alone.

\section{Discussion}
The SSA22-HCM1 galaxy is characterized by both a strong color break and strong
Ly$\alpha$ emission.  The high contrast of this object in the emission-line
bandpass can be seen in the narrowband 8185/105 \AA\ panel of Fig.~\ref{fig:1},
and in the extended appearance of SSA22-HCM1 in this band compared with the red
$Z$ and $I$ continuum bandpass images.  The strong continuum break is seen in
the SED (Fig.~\ref{fig:2}), with at least a factor 4.2 jump based on the
$(R-Z)$ color difference.  It is completely absent in the $B$-, $V$-, and most
significantly, $R$-band images.  For the $R$-band non-detection we use the
$2\sigma$ estimate of 27.1 ($AB$ magnitude) to set the lower limit of $(R-Z) >
1.6$ and estimate a minimum break strength.  The break may also be seen in the
spectral data, and in Fig.~\ref{fig:3} we show a composite 10800 s LRIS
spectrum.  The heavy lines show the averaged continuum values for the
wavelength regions sampled above and below the strong emission feature, and
while there is a significant positive signal above the line the spectrum is
consistent with there being zero flux at shorter wavelengths.  Fig.~\ref{fig:4}
shows the region of the 8190 \AA\ emission line in the unsmoothed summed
spectrum of all observations made with the $300\ell$/mm grating (18,000 s). The
line is unresolved at an instrumental resolution of 12.4 \AA\ FWHM,
corresponding to a velocity $\sigma < 200$ km s$^{-1}$,  Instrument problems
prevented acquisition of sufficient higher resolution observations to test for
line asymmetry.  Examination of the spectral data provides a rough consistency
check of the continuum and line properties.  Following Oke and Korycansky
\markcite{okekor} (1982), we measure the flux deficit parameter $$D_A =
\left\langle 1 - {{f_{\nu}({\rm observed})} \over {f_{\nu}({\rm continuum})}}
\right\rangle$$ using the rest-frame wavelength range $1050-1170$ \AA\ for
$f_{\nu}({\rm observed})$ and $1250-1370$ \AA\ for the red observed continuum.
These yield $f_{\nu} = 0.050 \pm 0.048 \mu$Jy and $f_{\nu} = 0.244 \pm 0.054
\mu$Jy, respectively.  This gives a formal ratio $D_A = 0.79$, similar to the
limit obtained with the $Z$- and $R$-band filters ($D_A > 0.76$), and again it
is the limits on the non-detection of the galaxy at shorter wavelengths that
constrains the lower limit on $D_A$.  The errors in these faint continuum
measurements based on spectroscopy are dominated by nightsky subtraction and
systematic errors, and are not Gaussian.  The filter photometry provides a more
robust and conservative determination of the continuum break.  However, the
emission line flux calculated from the spectroscopic data ($2.0 \pm 0.1 \times
10^{-17}$ ergs cm$^{-2}$ s$^{-1}$) and the continuum magnitudes
($AB_{\lambda\lambda 1050-1170,\,{\rm rest}} < 27.0$ and $AB_{\lambda\lambda
1250-1370,\,{\rm rest}} = 24.4 \pm 0.3$) agree with the $R$ and $Z$ continuum
magnitudes listed in Table~\ref{tbl:1}, and are consistent with a flat $f_{\nu}$
spectrum plus a break.  Identifying the line with Ly$\alpha$, we find the
redshift of $z=5.74$.  The remaining properties of the line are best determined
from the imaging data, and we find from these that it has an observed equivalent
width of 175 \AA\ and a flux of $1.8\times 10^{-17}$ erg cm$^2$ s$^{-1}$.

The object most similar to SSA22-HCM1 among those known in the current
literature is the $z=5.60$ galaxy, HDF 4-473.0 \markcite{weymann98}(Weymann
{et~al.}\ 1998), which shows both the strong continuum break between the F606W
and F814W (and infrared F110W and F160W bands at $J$ and $H$), and an emission
line of roughly comparable equivalent width and flux, corresponding to
Ly$\alpha$ at $z=5.60$.  The Ly$\alpha$ flux of SSA22-HCM1 is approximately
twice as strong, and the continuum above the break a factor of 3 times more
luminous than for this $z=5.60$ galaxy, using the near-IR magnitude of HDF
4-473.0, AB(F110W)=26.64, to approximate its continuum.  In comparison, while
the estimated continuum magnitudes above Ly$\alpha$ for the two
\markcite{spin98}Spinrad {et~al.}\ galaxies are comparable to that of SSA22-HCM1
(24.9 and 25.7 AB mags for 3-951.1 and 3-951.2 vs.\ 25.5, estimated from our
$Z$-band observations), neither of these systems shows any emission, while the
\markcite{dey98}Dey {et~al.}\ object, RD1, has much higher equivalent width (600
\AA), and a fainter continuum above Ly$\alpha$ (26.3 AB mags based on the much
more uncertain spectroscopic estimates).

\begin{inlinefigure}
\centerline{\epsfig{file=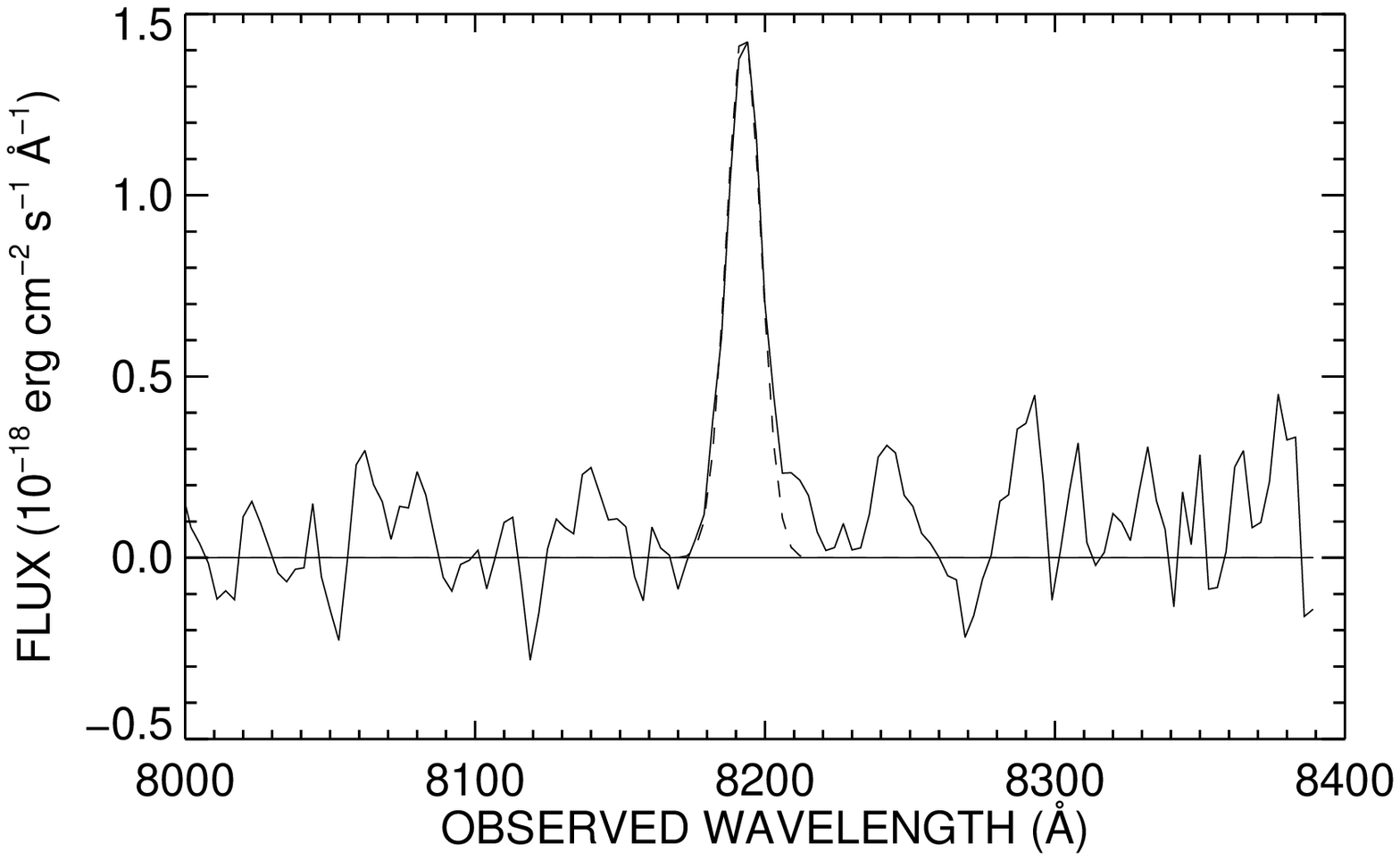,width=235pt}}
\vspace{-0.25cm}
\caption{\small Spectral region around the emission. The complete
summed LRIS observations from 18,000 s total exposure with the $300\ell$/mm
grating. The instrumental resolution (12.4 \AA\ FWHM, determined from
Gaussian profile fits to night sky lines) is overplotted with the dashed
curve.  The line is unresolved at a velocity $\sigma < 200$ km s$^{-1}$.}
\label{fig:4}
\end{inlinefigure}

Since resonant scattering enhances the effects of extinction, converting the
Ly$\alpha$ emission into a massive star formation rate entails large
uncertainties.  For the present calculation, we assume that extinction may be
neglected in computing the required massive star formation rates, which then
constitute a minimum estimate.  The maximum values obtainable from ionization
by a massive star population lie in the $100-200$ \AA\ range
\markcite{charl93,dvg93}(Charlot \& Fall 1993; {Valls-Gabaud} 1993).  Up to
roughly half the rest-frame emission may be eaten away by the Lyman alpha
forest, and applying such a correction factor would give values above 50 \AA,
as compared to the \markcite{weymann98}Weymann {et~al.}\ rest-frame equivalent
width (corrected for hydrogen absorption) of 90 \AA.  Assuming case B
recombination, we have $L($Ly$\alpha$) = 8.7\,$L($H$\alpha$)
\markcite{brock}(Brocklehurst 1971), which using \markcite{kenn83}{Kennicutt,
Jr.}'s \markcite{kenn83} (1983) translation of $\dot{M}$ from H$\alpha$
luminosity, gives $\dot{M}=(L($Ly$\alpha$)/$10^{42}$ ergs s$^{-1}$) $M_{\odot}$
yr$^{-1}$. For $H_0=65$ km s$^{-1}$ and $q_0=0.02$, the observed line flux of
$1.8\times 10^{-17}$ erg cm$^{-2}$ s$^{-1}$ translates into a star formation
rate of 19 \msun\ yr$^{-1}$. (For $q_0=0.125$, the line luminosity would be a
factor of 1.9 times lower; and for $q_0=0.5$, a factor of 6 times lower than
for the present calculation.)  These values, which are uncorrected for dust
extinction, are consistent with the maximum values seen in the estimates for
the systems up to $z\sim4.5$ in the Lyman break galaxies studied to $I(AB)<25$,
which have rest-frame equivalent widths up to 80 \AA\ \markcite{stei99}(Steidel
{et~al.}\ 1999).  For $H_0=65$ km s$^{-1}$ and $q_0=0.02$, the continuum
luminosity at rest-frame 1500\AA\ is $9.7\times 10^{-20}$ erg cm$^{-2}$
s$^{-1}$ \AA$^{-1}$, and corresponds to a star formation rate
\markcite{madau98}(e.g., Madau, Pozzetti, \& Dickinson 1998) of roughly 40
\msun\ yr$^{-1}$, consistent with the lower limit set by the emission line
calculation.

Because SSA22-HCM1 is so bright in $Z$ it is of interest to estimate the number
of potential high redshift ($z>5$) objects from ($R-Z$) color statistics using
a $Z<25.25$ sample.  (Although this $Z$ band criterion is not sufficiently deep
to include SSA22-HCM1, the $z=5.60$ galaxy HDF 4-473.0
\markcite{weymann98}(Weymann {et~al.}\ 1998), or the $z=5.34$ emission-line
object \markcite{dey98}($I(AB)=26.1$; Dey {et~al.}\ 1998), the threshold is only
a few tenths of a magnitude from reaching the brighter two of these objects.)\
Line blanketing from the intergalactic Lyman alpha forest quenches the flux
below redshifted Lyman alpha by more than an order of magnitude, based on
extrapolations from forest simulations \markcite{zhang97}(e.g., Zhang {et~al.}\
1997).  For the SSA22 LRIS field over a 32 arcmin$^2$ region there are only
four objects with $(R-Z)> 2.75$ and $Z(AB) < 25.25$, with typical
$Z(AB)\sim25$, with a similar number in an LRIS field crossing the HDF, for
which the effective $Z$ exposure is roughly twice as deep.  Such objects may be
$z>5$ galaxies or highly reddened objects; the color boundary lies marginally
above the range of possible M star contaminants, which might be distinguished
by compactness criteria. In the HDF the ($R-Z$) method recovers the combined
$V$-dropout objects, HDF 3-951.1 and HDF 3-951.2, which
\markcite{spin98}(Spinrad {et~al.}\ 1998) identify as a $z\sim5.34$ galaxy pair
based on a spectral break (estimated redshift and discontinuity of $> 8 $ based
on the LRIS continuum spectrum, with no emission features).  Our $Z$ and $R$
imaging gives $(R-Z) \sim 3.1$ for this object, similar to the colors of the
Ly$\alpha$ bright objects.  Thus, it appears that very deep $Z$ imaging can
reveal the very brightest members of the redshift 5 population and represents a
viable alternative way to approach the problem.

\acknowledgements

It is a pleasure to thank Alan Stockton for many helpful discussions on filters
and optimal strategies for studying high-$z$ galaxies, and specifically for the
use of his RG850 filter for this program.  We thank T.\ Bida, R.\ Goodrich,
T.\ Chelminiak, R.\ Quick, and W.\ Wack for their assistance in obtaining the
observations, which would not have been possible without Bev Oke and Judy
Cohen's LRIS spectrograph; and we thank Bill Mason for heroic hardware
support.  This work was supported in part by the State of Hawaii and by NASA
grants GO-5922.01-94A, GO-6626.01-95A, and GO-7266.01-96A from Space Telescope
Science Institute, which is operated by AURA, Inc., under NASA contract NAS
5-26555.  E.M.H.\ and L.L.C.\ thank the IoA, Cambridge, for its hospitality
during the writing of this paper.  R.G.M. thanks the Royal Society for
support.


\end{document}